# The Structural Phase Transition of the Relaxor Ferroelectric 68%PbMg$_{1/3}$Nb$_{2/3}$O$_3$-32%PbTiO$_3$

S. N. Gvasaliya[1,2], B. Roessli[1], R. A. Cowley[3], S. Kojima[4] and S.G. Lushnikov[2]

[1] Laboratory for Neutron Science ETHZ and Paul Scherrer Institut, CH-5232, Villigen PSI, Switzerland
[2] Ioffe Physical Technical Institute, 26 Politeknicheskaya, 194021, St. Petersburg, Russia
[3] Clarendon Laboratory, Department of Physics, Oxford University, Parks Road, Oxford, OX1 3PU, UK
[4] Institute of Materials Science, University of Tsukuba, Tsukuba, Ibaraki 305-8573, Japan

**Abstract**
Neutron scattering techniques have been used to study the relaxor ferroelectric 0.68PbMg$_{1/3}$Nb$_{2/3}$O$_3$-0.32PbTiO$_3$ denoted in this paper as 0.68PMN-0.32PT. On cooling, these relaxor ferroelectrics have a long-range ordered ferroelectric phase and the composition is close to that at which the ferroelectric structure changes from rhombohedral to tetragonal. It was found that above the Burns temperature of about 600K, the transverse optic mode and the transverse acoustic mode are strongly coupled and a model was used to describe this coupling that gave similar parameters to those obtained for the coupling in PMN. Below the Burns temperature additional quasi-elastic scattering was found which increased in intensity as the sample was cooled down to the ferroelectric transition temperature but then decreased in intensity. This behaviour is similar to that found in PMN. This scattering is associated with the dynamic polar nano-regions that occur below the Burns temperature. In addition to this scattering a strictly elastic resolution limited peak was observed that was much weaker than the corresponding peak in pure PMN and which decreased in intensity on cooling below the ferroelectric phase whereas for PMN, which does not have a long-range ordered ferroelectric phase, the intensity of this component increased monotonically as the sample was cooled. The results of our study are compared with the recent measurements of Stock et al. [PRB **73** 064107] who studied 0.4PMN-0.6PT. In this material the ferroelectric transition is at 550 K and there is little difference between the Burns temperature and the ferroelectric transition temperature while only weak diffuse scattering was observed by thermal neutron scattering. This shows that quasi-elastic or central peak observed both in PMN and in the present study result from scattering by dynamic and static nano-regions. The results are qualitatively consistent with the random field model developed to describe the scattering from PMN.

Communicating Author: S. Gvasaliya

PACS: 77.80-e, 61.12-q, 63-50+x, 64.60-i.

## 1. Introduction

The relaxor ferroelectrics have been known for over 50 years but their properties are not yet understood and continue to be the subject of many experiments and theoretical suggestions [1-5]. Much of the effort has been concerned with experiments on either PMN (PbMg$_{1/3}$Nb$_{2/3}$O$_3$) or on PZN (PbZn$_{1/3}$Nb$_{2/3}$O$_3$) and with these materials doped with PT (PbTiO$_3$). The crystal structure of PMN is cubic in the absence of an applied electric field down to the lowest temperature [6]. Below the Burns temperature, however, which is about 620K in PMN, optical measurements [7] suggest that the crystal breaks up into polar nano-regions which are locally ferroelectric while the transverse optic mode at long wavelength becomes over-damped. X-ray and neutron scattering measurements show strong diffuse scattering [6, 8], both quasi-elastic and strictly elastic [9]. In addition, neutron scattering studies indicate that the transverse acoustic (TA) and the lowest transverse optic (TO) branches are coupled [10, 11]. Both Stock et al [12] and ourselves [11] have suggested that these properties can be understood in terms of random fields in the cubic crystal. At high temperatures but below the Burns temperature, 620 K, there are dynamic nano-regions produced by the random fields. In this case the random fields have isotropic symmetry and as suggested by Imry and Ma [13] there is then no long range ordered phase. At lower temperatures, below 370 K, the fields develop a cubic anisotropy and the system has an increasing number of static nano-regions as the crystal if further cooled. The system has no long range order except in the presence of an electric field which is similar to the behaviour found for Ising random fields in magnetic systems [14]

Lead titanate, PT, is a more conventional ferroelectric crystal with a soft mode and a transition to a long-range ordered structure occurring below 770K [15]. The low temperature phase is tetragonal and similar to that occurring in ferroelectric BaTiO$_3$ just below its Curie point. Mixed crystals can be grown for all concentrations 1-x of PMN and x of PT and the phase diagram has been measured in detail [16-20] and is shown in fig.1. For small x, small amounts of PT, there is a transition to a long-range ordered ferroelectric state, although as commented above this transition does not occur in pure PMN. When the concentration x is increased to about 0.32 the ferroelectric transition temperature increases from around 250K to about 400K and throughout most of this range the long-range ordered ferroelectric phase is rhombohedral. There is some evidence of distortions from the rhombohedral structure, for x>0.28, but, by and large, the structure of this phase is clear. When the concentration of PMN is further increased above x=0.5 the transition temperature rises and reaches 770K when x=1 while the long-range ordered ferroelectric phase occurring below the transition temperature is tetragonal. Between the rhombohedral phase and the tetragonal phase the measurements are difficult because the measurements are often made with thin films and the effect of strains from the substrates and weak electric fields can be considerable. Nevertheless the structure of this phase is monoclinic and birefringence measurements suggest that there is a first order transition between a phase with Cm symmetry and one with Pm symmetry. Both phases seem to have continuous phase transitions to the rhombohedral and tetragonal phases.

The structure of these phases can be understood qualitatively by using the Landau theory of phase transitions. The free energy is expanded in powers of the spontaneous polarisation **P** = (P$_x$, P$_y$, P$_z$) as [21]:

$$F = A(T - T_c)P^2 + BP^4 + C(P_x^4 + P_y^4 + P_z^4) + \ldots (1)$$

where the ferroelectric transition occurs when T=$T_c$ and $A$, $B$, and $C$ are constants. The first two terms in this expansion are isotropic terms and so do not influence the direction of the spontaneous polarisation. The third term is anisotropic and in conventional ferroelectrics the direction of the polarisation is dependent on the sign of the coefficient $C$ or in some cases on higher order terms. Low order perturbation theory shows that if B>0 and $C$ is positive the minimum of the free energy occurs if the spontaneous polarisation is along a [1,1,1] direction giving a rhombohedral structure, while if $C$ is negative the minimum occurs when the polarisation is along the [1,0,0] direction giving a tetragonal structure. It is then natural to describe the phase diagram of PMN-PT by assuming that $C$ is positive for x<0.32 and negative for x>0.32 at least close to the onset of ferroelectricity. Of course in these diffuse ferroelectrics the properties are more complicated because of the random electric fields and higher order terms in the expansion of the free energy as described at least in part by Vanderbilt and Cohen [22].

Nevertheless similar behaviour occurs in the mixed rare earth hexagonal systems, such as $Er_xHo_{1-x}$ and $Tm_xHo_{1-x}$. In these materials the anisotropy is provided by the quadratic terms in the free energy of the form $CP_z^2$ and C varies from negative to positive as the concentration changes. When C is negative the ordered structure is aligned along the c axis (Er or Tm) while if C is positive the ordered structure is aligned in the xy plane (Ho). The concentration at which C and the isotopic quadratic term are zero is a multi-critical point and a detailed analysis and experiment then show that 5 different phases all meet at this point [23,24].

Although there have been many studies of the structure of the mixed crystals there have been very few studies of the dynamical properties and of the phonon spectra. Recently there has been a neutron and x-ray scattering study by Stock et al [25] of samples with x=0.6, well into the tetragonal part of the phase diagram. The results showed that the diffuse scattering was absent or at least much weaker in this sample than in PMN and it was suggested that the polar nano-regions found in relaxor ferroelectrics were not present. Nevertheless aspects of the scattering, such as the waterfall effect, were observed and interpreted as arising from random fields.

In this paper we report on high resolution neutron scattering measurements on a sample with x=0.32 close to the concentration where the averaged cubic anisotropy is expected to be very small. The properties in this region of the phase diagram have been studied in detail because the samples are useful for piezoelectric applications. This is because the ordered ferroelectric moments can be rotated very easily by applying stresses to the materials and as a result the piezoelectric constants are unusually large and have uses in applications. The measurements were performed with a similar configuration to that used to study the dynamical properties of PMN [11]. The details of the sample and of the resolution of the neutron spectrometer are given in section 2. In section 3 we describe the measurements made and, as also found in PMN, there is a strong coupling between the transverse optic mode and transverse acoustic mode. In a final section we discuss our results and compare them with the results on PMN and with those obtained by Stock et al with x=0.6 [25].

## 2. Experimental Details

A high quality crystal of 0.68PMN-0.32PT was grown by TRS ceramics. The mosaic spread is less than 0.5°. The sample of cubic shape with a size of 1 cm³ was aligned in the (hhl) scattering plane and placed in a niobium holder which could be mounted inside a furnace.

The measurements were performed with the TASP [26] triple axis spectrometer located at the spallation neutron source, SINQ [27], of the Paul Scherrer Institut in Switzerland. The spectrometer was configured with both a pyrolytic graphite (PG) monochromator and analyser and operated so that the scattered energy was held constant. For most of the experiments the collimation from source to detector was $0.63^0$-$0.67^0$-$0.67^0$-$1.33^0$ horizontally and the final scattered wavelength was 3.83 Å. A PG filter was used in the scattered beam to eliminate higher order neutrons. The energy resolution at zero energy transfer was 0.2 meV, FWHM, and the wave-vector resolution approximately 0.012 Å⁻¹, 0.015 Å⁻¹ and 0.1 Å⁻¹ parallel to the wave-vector transfer, perpendicular but in the horizontal plane and vertically respectively.

Most of the data was obtained by scanning the spectrometer so that:
   a.  the energy or wave-vector was held constant in either the (1,1,0) or (2,2,0) Brillouin zones while the temperature was held at 700 K.
   b.  the wave-vector or energy were held constant in the (1,1,0) Brillouin zones while the temperature was kept at either 430 K or 475 K.
   c.  the wave-vector was held constant at either (1,1,0.75) or (2,2,0.75) while the temperatures was varied between 330 K and 700 K.

In this description, and henceforth in this article, the wave-vector transfers are given in reciprocal lattice units (rlu), $2\pi/a$.

## 3. Experimental Results

### 3.1    Phase transitions

The Bragg scattering from the structure of the sample of 0.68PMN-0.32PT were recorded as a function of temperature so as to determine whether that were consistent with the phase diagram given in fig. 1. The results for the temperature dependence of the scattering from the (002) Bragg peak are shown in fig. 2. There is a change in the slope of the curve at about 425K which agrees with the temperature of the continuous transition to the ferroelectric phase with a structure which is at least close to being tetragonal. There is then a rapid increase in the intensity between 370K and 380K which is consistent with the transition from the Pm phase to the Cm phase and is possibly of first order although slightly smeared by any concentration gradient across the crystal. The temperature is slightly larger than that found by Shuvaeva et al. [19], but this is probably due to the different origin of the samples.

### 3.2    Measurements above the Burns Temperature at 700K

The neutron scattering was measured near the (1,1,0) and (2,2,0) reciprocal lattice points. As shown in fig. 3, the inelastic scattering near the (1,1,0) lattice point was dominated by the transverse acoustic mode with only a small amount of scattering that could be ascribed to the transverse optic mode. The relative intensity of the inelastic scattering steadily decreases compared with the elastic scattering as the wave-vector transfer increases. In contrast near the (2,2,0) reciprocal lattice point the transverse optic mode has the more intensity, as shown in fig. 4, which shows the results for both constant wave-vector scans and for constant energy scans.

These results are similar to those for PMN at high temperatures and show that there is no temperature dependent quasi-elastic scattering at 700K in 0.68PMN-0.32PT. Consistently with our results on PMN we then assume that the temperature in this material was above its Burns temperature. Unfortunately we are not aware of any determination of the Burns temperature in the doped material. The scattering profiles have been fitted to two coupled modes for each scan. Using the same notation as previously [11] the energy of the transverse acoustic mode was parameterized as:

$$\omega_{TA} = d \sin(\pi q) \qquad (2)$$

the energy of the transverse optic mode as:

$$\omega_{TO}^2 = \omega_0^2 + c \sin^2(\pi q) \qquad (3)$$

with the coupling between the two modes having the form:

$$\Delta = \Delta_{12} \sin^2(\pi q) \qquad (4).$$

The damping of the transverse acoustic mode is given by:

$$\gamma_{TA} = D_{TA} \sin^2(\pi q) \qquad (5)$$

while for the optic mode the damping was chosen for each wave vector. The coupled mode model was used to calculate the susceptibility and then the neutron scattering cross section as described in our earlier paper on PMN [11]. The parameters were fitted to the experimental results and the energies of the modes are shown in fig. 5 with the parameters $d = 11.4 \pm 0.4$ meV, $\omega_0 = 2.7 \pm 0.2$ meV, and $c = 165.4 \pm 9.8$ meV$^2$. The coupling between the modes is given by $\Delta_{12} = 117.3 \pm 4.1$ meV$^2$. These are slightly larger than the parameters in PMN with the increase varying between 28% for $\Delta_{12}$ and 12% for d. The value of the imaginary part of the coupling term was found to be similar to PMN and hence fixed at the value $\gamma_{12}$=1.5 meV. The damping of the transverse optic and transverse acoustic modes are shown in fig. 6. The results for the transverse acoustic mode are a reasonable smooth curve with $D_{TA}$=3.5$\pm$0.2 meV. This value is lower by 17% than the value of the damping observed in PMN. The damping of the transverse optic mode shows considerable scatter but has a value that is about 1.5 meV. The indication is that the transverse optic mode at the zone centre does not completely soften as found in most other relaxor ferroelectrics.

### 3.3 The Temperature Dependence of the Spectrum and Quasi-elastic Scattering

The neutron scattering at wave-vector transfers of (1,1,0.075) and (2,2,0.075) were measured at a range of temperatures between 330 K and 700 K, as shown in fig. 7. Fits were then made using the coupled phonon model described in the previous section with the addition of both a quasi-elastic peak and a strictly elastic peak. There was little change in the inelastic parameters as shown in fig. 8 for the dispersion and structure factor of the transverse acoustic mode. Both decreased on cooling from 700 K to about 450 K and then increased slightly as the temperature was further cooled. The broad minimum in the curves is, within error, consistent with the temperature of the ferroelectric phase transition. Note that the corresponding values obtained directly from data taken near the (2,2,0) Bragg reflection show temperature independent behaviour. The damping of the TO phonon increases gradually on approaching the phase transition as shown in fig. 9.

There is, however, considerable change in the scattering in the region around the elastic peak. As shown in fig. 7 there is an increasing intensity of quasi-elastic scattering as the temperature is reduced from 700K to 430K. On further cooling to 330K the quasi-elastic scattering decreases as shown in fig. 10 which shows the temperature dependence of the quasi-elastic scattering at the wave-vector transfer of (1,1,0.075) that is compared with the temperature dependence of the dielectric susceptibility. Both have a peak centred at a temperature of 430 K although the dielectric permeability is sharper than the maximum of the static susceptibility $\chi(0,T)$ measured from the quasi-elastic scattering. Possibly this is because the quasi-elastic scattering is measured at a non-zero value of the wave-vector transfer, namely (1,1,0.075), while the dielectric susceptibility is not. Similar results were deduced from the scattering at (2,2,0.075). The energy width of the scattering steadily decreases from about 1 meV just below 700K to a minimum energy width of just below 0.5 meV at 430 K. On further cooling the energy width tends to increase in width as shown in fig. 11. These results are similar to those obtained in PMN, except that the energy width of the scattering steadily decreases on cooling in PMN in contrast to the increase observed in the doped material below 420 K.

A detailed study of the shape of the quasi-elastic scattering was made at both 430 K and 475 K by measuring a series of scans with the wave-vector transfer along the (1,1,q) direction with component q varying between 0.035 and 0.125 rlu. The results of the fitting to the coupled mode model with an additional quasi-elastic component are given in figs. 12, 13 and 14. Figure 12 shows the wave-vector dependence of the quasi-elastic scattering intensity at 475K and it is clearly peaked at small wave vectors and a fit to a Lorentzian curve gives an inverse correlation length of about 0.039 rlu. Figure 13 shows the width compared with that observed in PMN. The magnitudes are similar although there is less temperature dependence for the present sample than for PMN. Figure 14 shows the increasing width in the quasi-elastic energy when the wave vector is increased. The results are consistent with fits to:

$$\Gamma(q) = \Gamma_0 + D_{QE}q^2 \qquad (6)$$

with $\Gamma_0 = 0.19 \pm 0.09$ and $0.26 \pm 0.06$ meV at 430 K and 475 K respectively while $D_{QE} = 64 \pm 23$ and $54 \pm 12$ mev/q$^2$.

### 3.4 Central Peak Intensity

The spectra shown in fig.7 also show that there is considerable central peak intensity, with $\omega = 0$, and resolution limited. The temperature dependence of the scattering intensity measured at a wave-vector transfer of (1,1,0.75) is shown in fig. 15. The intensity is almost independent of temperature, above a temperature of 550 K, at a value presumably given by the defect scattering and the incoherent scattering. The scattering then increases on cooling to a temperature of about 400 K. The intensity then decreases on further cooling. Unfortunately we were unable to obtain a precise wave-vector dependence of this scattering due to the relatively low intensity of the central peak as compared to both the quasi-elastic scattering and the Bragg peak.

## 4. Discussion

PMN doped with 32% of PT has been studied with neutron inelastic scattering techniques. The results show that at high temperature, well above the Burns temperature, there is a strong coupling between the TA mode and the lowest frequency TO mode that can be analysed with the same formalism as was applied to PMN. On cooling the sample below the Burns temperature the results of the experiments show that in addition to transverse acoustic and optic phonons both a quasi-elastic peak and a resolution limited central peak are observed in the scattering. The quasi-elastic scattering peak has a broad maximum at approximately the temperature of the ferroelectric phase transition where both the damping and the inverse correlation length have minima. The intensity of the central peak also has a weak maximum at about the same temperature. At the temperature where the quasi-elastic scattering appears the damping of the TO phonon starts to increase and the stiffness of the TO branch indicates a slight maximum. This may be a consequence of their scattering by the dynamic nano-regions.

Above the ferroelectric phase transition temperature these results are similar to those obtained from PMN except for the fact that the central peak is weaker in intensity and persists for temperatures well above the maximum of the dielectric susceptibility. Below the ferroelectric phase transition temperature both the quasi-elastic peak and the central peak decrease in intensity whereas for PMN the central peak continued to increase in intensity down to the lowest temperature [11]. This is consistent with the structure of PMN consisting of a large number of randomly oriented polar nano-regions whereas in PMN doped with 32% of PT the polar nano-regions all tend to become aligned and crystal is then a ferroelectric because the nano-regions contribute to the uniform ferroelectric moment. This is possibly surprising because this doped sample is very close to the concentration at which the cubic anisotropy is very small and we would expect there was little preference for any particular domain orientation. It must however be the case that the tendency to produce long range order is larger than the tendency to produce disorder and static frozen polar nano-regions. In terms of the random field model we used to explain our PMN results [11], the random fields do not produce long range order but produce dynamic polar nano-regions and these are observed to give rise to the quasi-elastic scattering. We then suggested that in PMN there was a transition to a state which was cubic in symmetry rather than having isotropic symmetry and in which the micro-domains became static. This was inferred from the temperature dependence of the central peak for which the intensity was large below the temperature of the maximum of the susceptibility of the quasi-elastic scattering and further the central peak was approximately constant below ~150 K where the quasi-elastic component disappeared. Much the same behaviour seems to occur in the doped sample except that static long-range ordered ferroelectricity replaces the static random cubic phase in PMN. This is shown by the increase in the intensity of the central peak when approaching $T_c$ from the paraelectric state and by the decrease of this scattering below a temperature of 400 K.

It is of interest to compare our results with those of Stock et al [25] who have studied with thermal neutron scattering a sample with 60% of PT and 40% of PMN. For these concentrations the sample is paraelectric above a temperature of 550 K and on cooling distorts to a long-range ordered tetragonal ferroelectric structure. In this material the damping of the transverse acoustic phonons was found to be small. Nevertheless the lowest transverse optic mode was soft below Tc and exhibited the waterfall effect. There was no quasi-elastic scattering measured which could be identified with polar nano-regions. On the other hand weak diffuse scattering was observed. We conclude that in this material there is less disorder while the ferroelectric transition temperature has increased. There is then a much smaller temperature region for which the polar nano-regions can be observed if at all. This picture is consistent with recent X-ray results in the classical ferroelectric with a soft mode like $PbTiO_3$ where no diffuse scattering was observed [28].

In conclusion we consider that our result on 0.68PMN-0.32PT taken together with those obtained on PMN [11] and on 0.4PMN-0.6PT [25] enable us to have a consistent phenomenological picture. At the PMN end of the phase diagram there is a strong coupling between the optic and acoustic transverse modes and this gives rise to the different line-shapes of the scattering in different Brillouin zones. At low energies and below the Burns temperature there is quasi-elastic scattering from the polar nano-regions which are in dynamic motion as predicted by the isotropic random field model [13]. At lower temperatures either there is a ferroelectric transition to long range order for concentrations near 0.32 of PT or the system fails to obtain long range order and has a structure consisting of static polar nano-regions as found for PMN. This may be associated with a change from isotropic symmetry to cubic symmetry and then possibly for the ferroelectric materials a consequent change to a distorted ferroelectric phase.

**Acknowledgements**

The experiments were performed at the spallation neutron source SINQ, Paul Scherrer Institut, Villigen (Switzerland). This research was partially supported by the Grant-in-Aid for Scientific Research (A), 16204032 under MEXT, Japan and by RFBR grant 05-02-17822.

**Figure Captions**

Fig. 1   The phase diagram of (1-x)PMN-xPT showing the symmetries of the different phases. The cubic phase is denoted C, the rhombohedral and tetragonal phases are R and T and between them are monoclinic phases Cm and Pm while the symmetry of the phase I' is still uncertain from ref. 19.

Fig 2.   The temperature dependence of the integrated intensity of the (002) Bragg peak of a 0.68PMN-0.32PT single crystal. Structural phase transitions can be identified for temperatures of 425 K and 375 K.

Fig 3.   Representative constant Q scans in the vicinity of the (1,1,0) Bragg peak at a temperature of 700 K. The fits are described in the text and shown by solid lines.

Fig 4.   Representative constant Q scans and constant energy scans with E=4 and 7 meV, in the vicinity of the (220) Bragg peak at a temperature of 700 K. The solid lines are the results of fits that are described in the text.

Fig. 5.   The dispersion curves for the TO and TA phonons propagating along the [001] direction. The two inner curves give the uncoupled energies of the modes while the outer two curves give the energies of the coupled modes.

Fig. 6.   The results of the fitting for the damping of the TA and TO modes from const-Q scans. The TA mode is given by the smooth curve and the open points the damping of the optic mode.

Fig. 7.   The temperature of the intensity obtained from experiments in which the wave-vector transfer was held fixed at (1,1,0.75). The fits are described in the text.

Fig. 8.   The temperature dependence of the slope (stiffness) and structure factor of the TA mode as measured near the (1,1,0) Bragg reflection. Note that the corresponding values obtained directly from data taken near the (2,2,0) Bragg reflection show temperature independent behaviour while the fractional changes for the (1,1,0.75) data are also small..

Fig. 9.   The temperature dependence of the damping of the TO phonon as measured near the (2,2,0) Bragg reflection. The inset shows the temperature dependence of the stiffness *c* of the TO phonon as deduced from the data at the (2,2,0.075) position.

Fig. 10. The temperature dependence of the susceptibility of the quasi-elastic scattering as measured at (1,1,0.75) compared with the dielectric susceptibility [20].

Fig. 11. The temperature dependence of the damping of the quasi-elastic scattering as measured at (1,1,0.75). The fits are to two linear power laws.

Fig. 12. The wave-vector dependence of the susceptibility of the quasi-elastic scattering measured close to the (110) Bragg reflection.

Fig. 13. The width in the reciprocal space of the quasi-elastic scattering as a function of temperature. The measurements were made near the (110) Bragg reflection and the results for 0.68PMN-0.32PT are compared with those from PMN.

Fig. 14. The wave-vector dependence of the damping of the quasi-elastic scattering measured in the vicinity of the (110) Bragg reflection.

Fig.15. The temperature dependence of the central peak (energy resolution limited peak) as a function of temperature. The results are taken from the data at a wave vector transfer of (1,1,0.75).

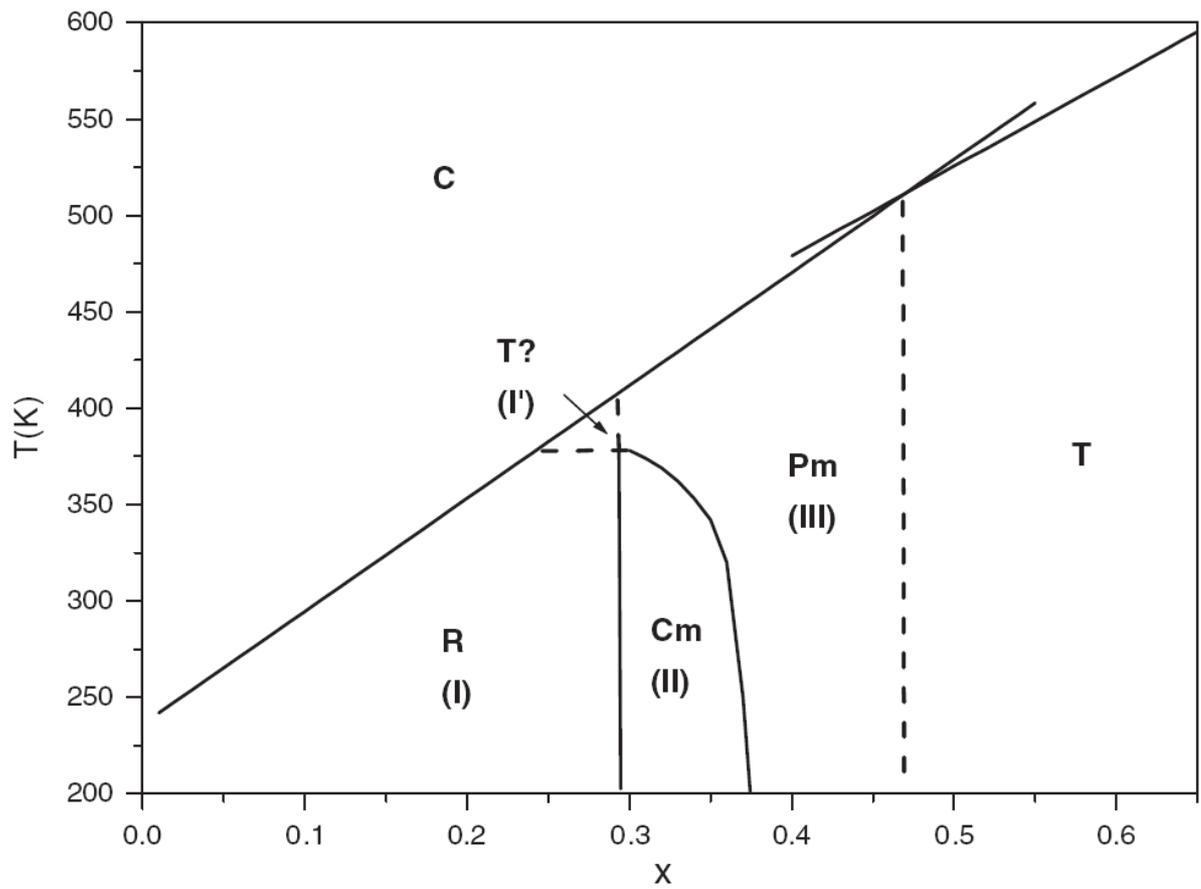

Fig. 1  The phase diagram of (1-x)PMN-xPT showing the symmetries of the different phases. The cubic phase is denoted C, the rhombohedral and tetragonal phases are R and T and between them are monoclinic phases Cm and Pm while the symmetry of the phase I' is still uncertain from ref. 19.

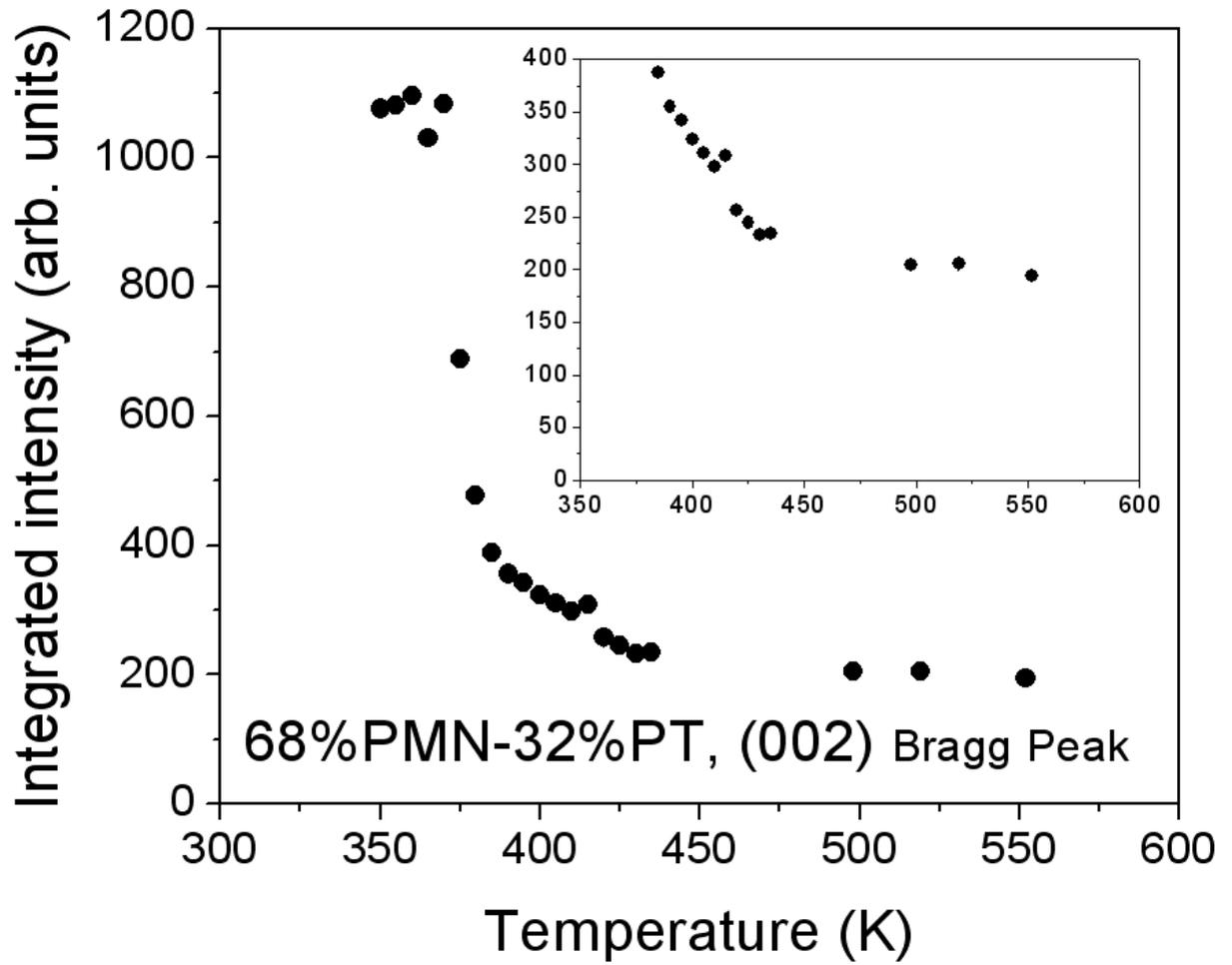

Fig 2. The temperature dependence of the integrated intensity of the (002) Bragg peak of a 0.68PMN-0.32PT single crystal. Structural phase transitions can be identified for temperatures of 425 K and 375 K.

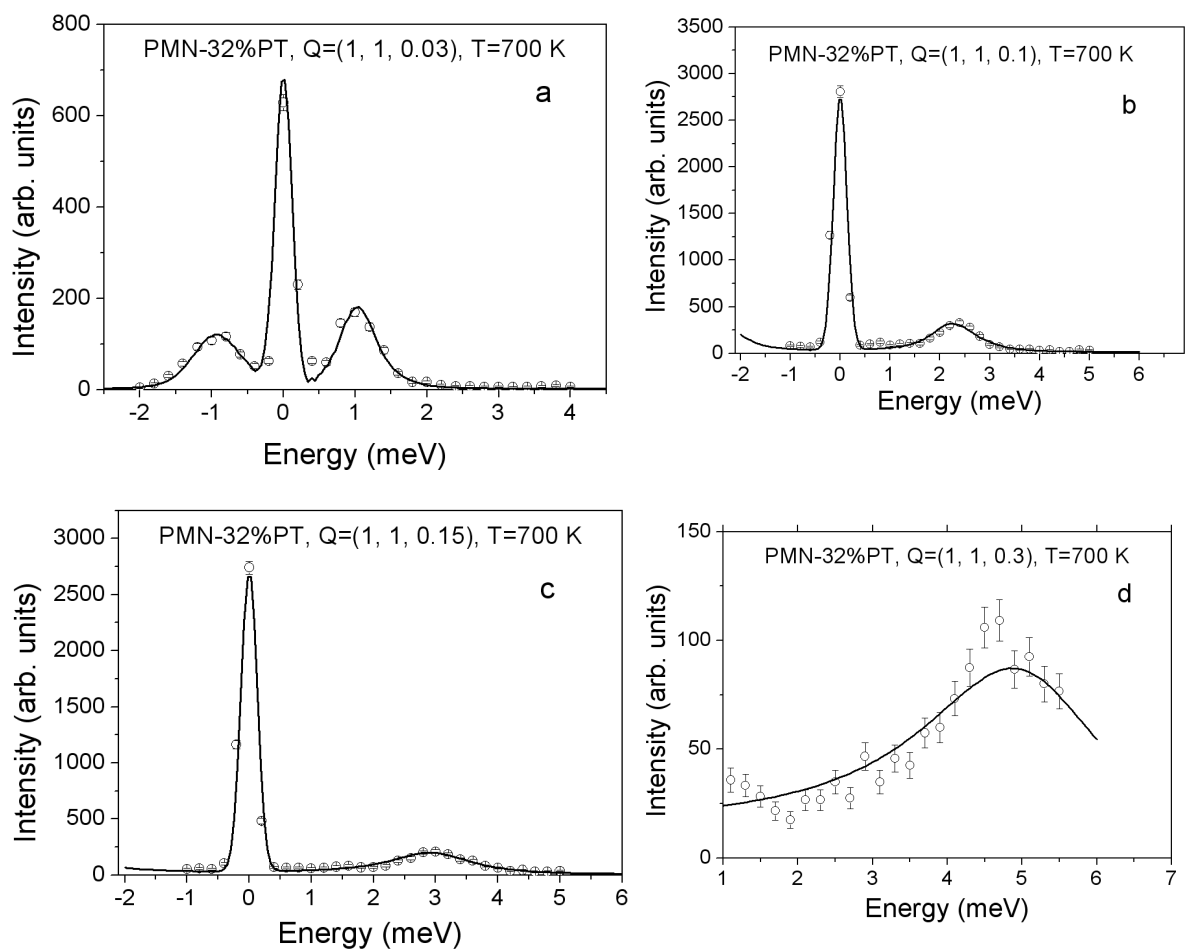

Fig 3. Representative constant Q scans in the vicinity of the (1,1,0) Bragg peak at a temperature of 700 K. The fits are described in the text and shown by solid lines.

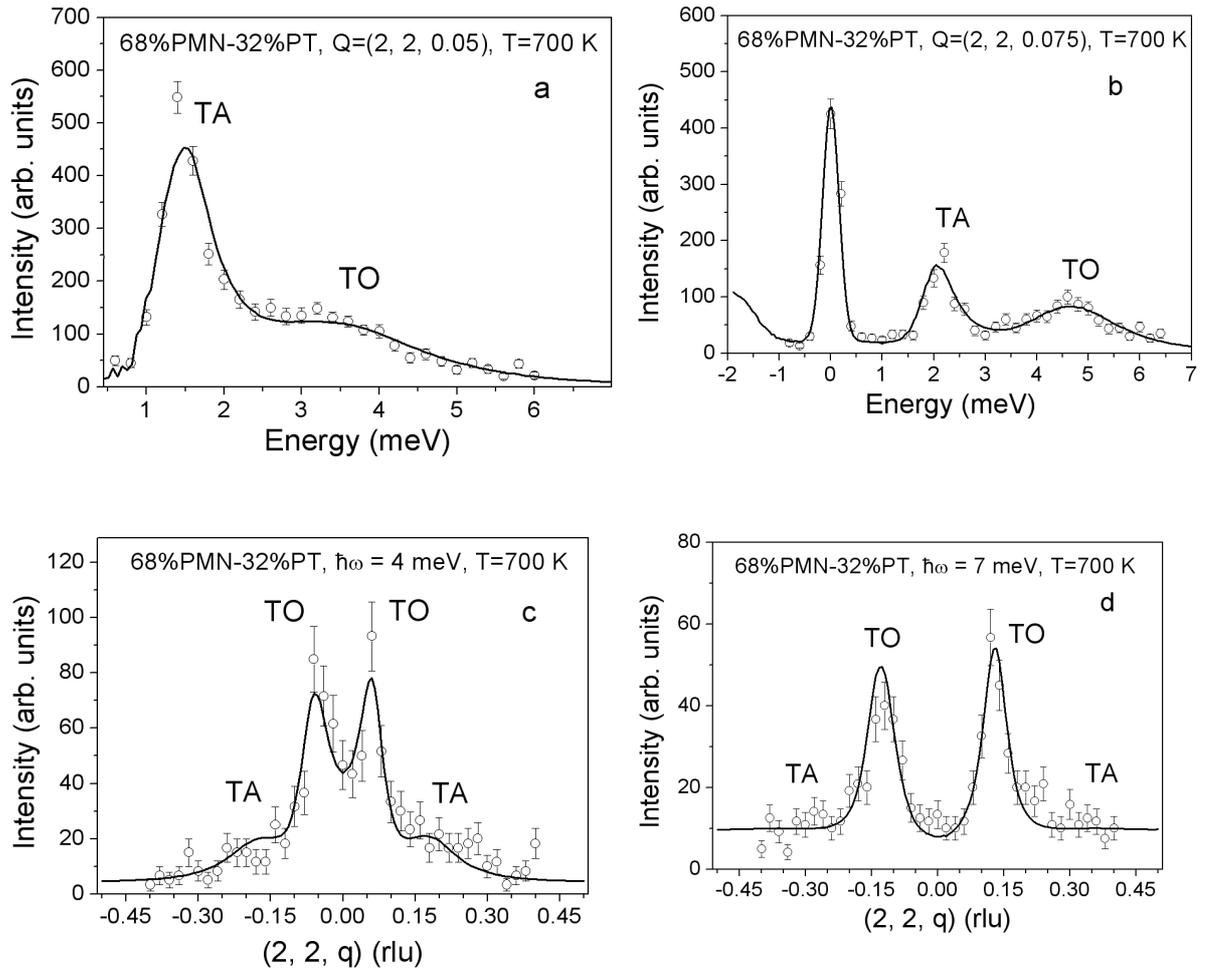

Fig 4. Representative constant Q scans and constant energy scans with E=4 and 7 meV, in the vicinity of the (220) Bragg peak at a temperature of 700 K. The solid lines are the results of fits that are described in the text.

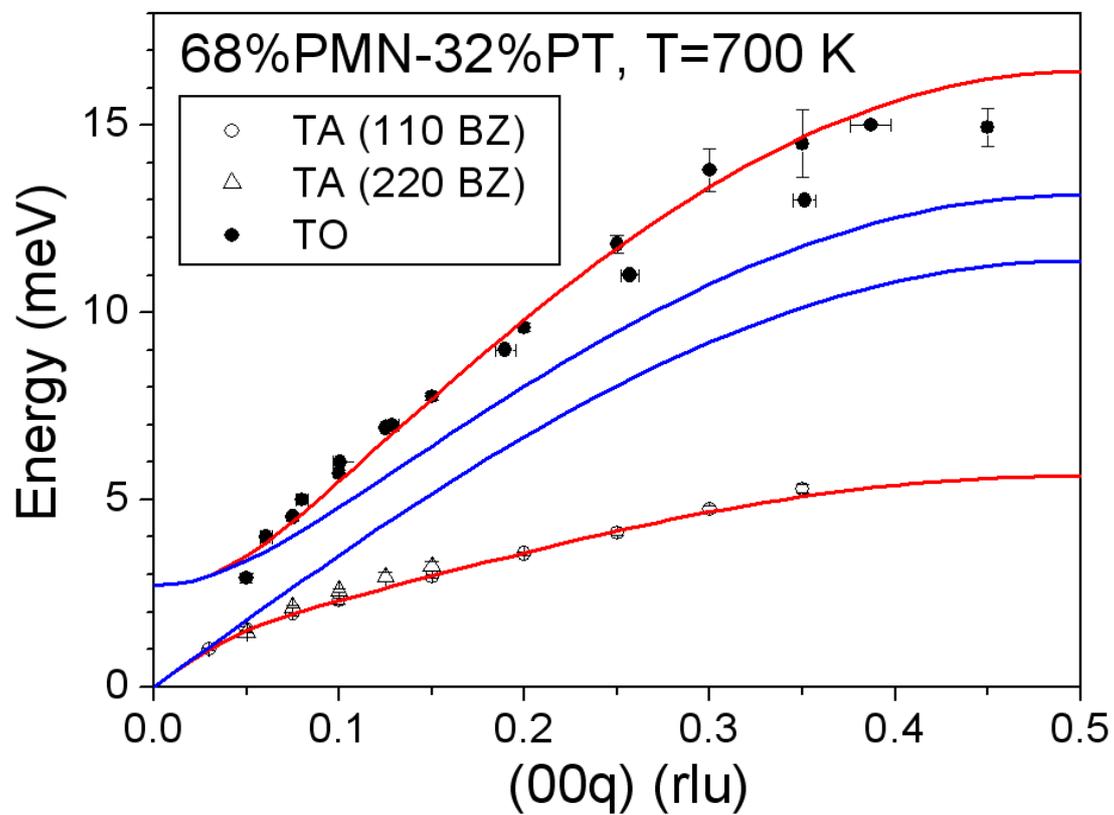

Fig. 5. The dispersion curves for the TO and TA phonons propagating along the [001] direction. The two inner curves give the uncoupled energies of the modes while the outer two curves give the energies of the coupled modes.

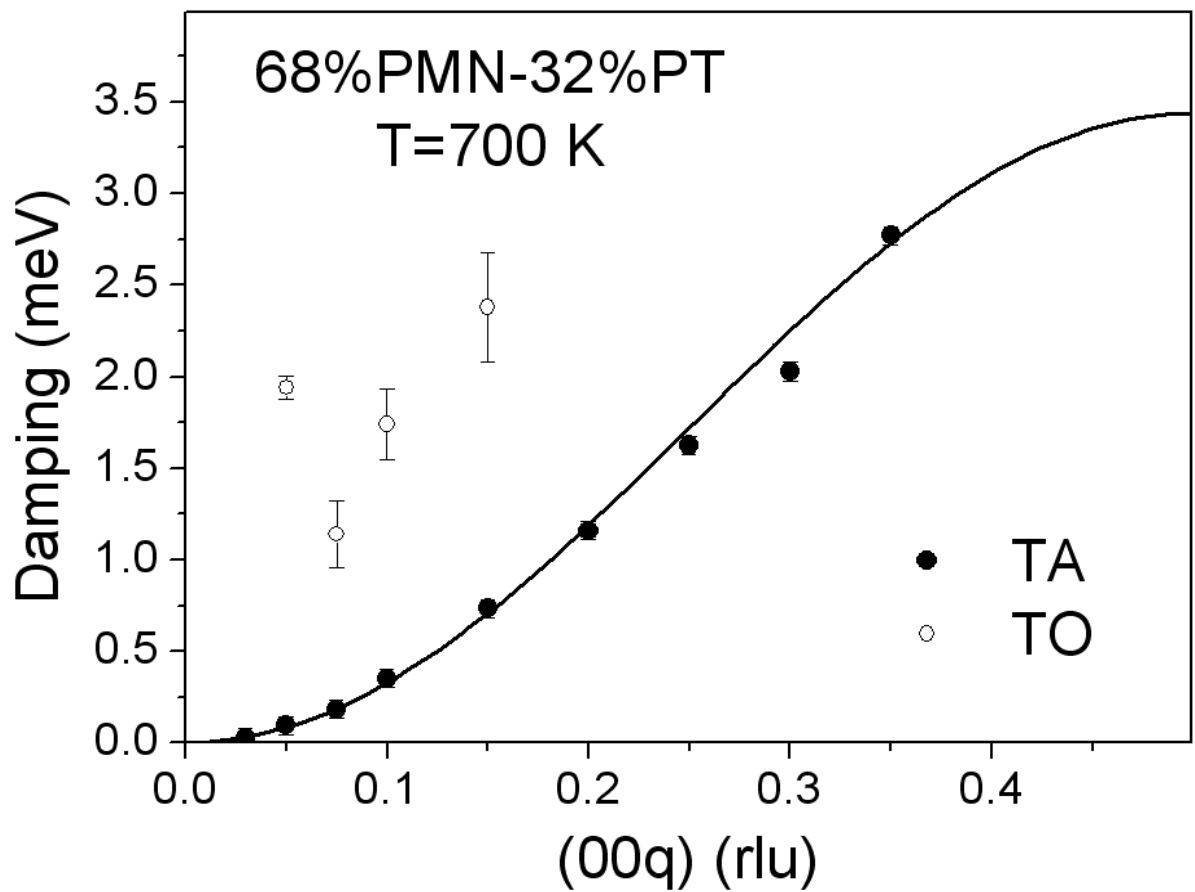

Fig. 6. The results of the fitting for the damping of the TA and TO modes from const-Q scans. The TA mode is given by the smooth curve and the open points the damping of the optic mode.

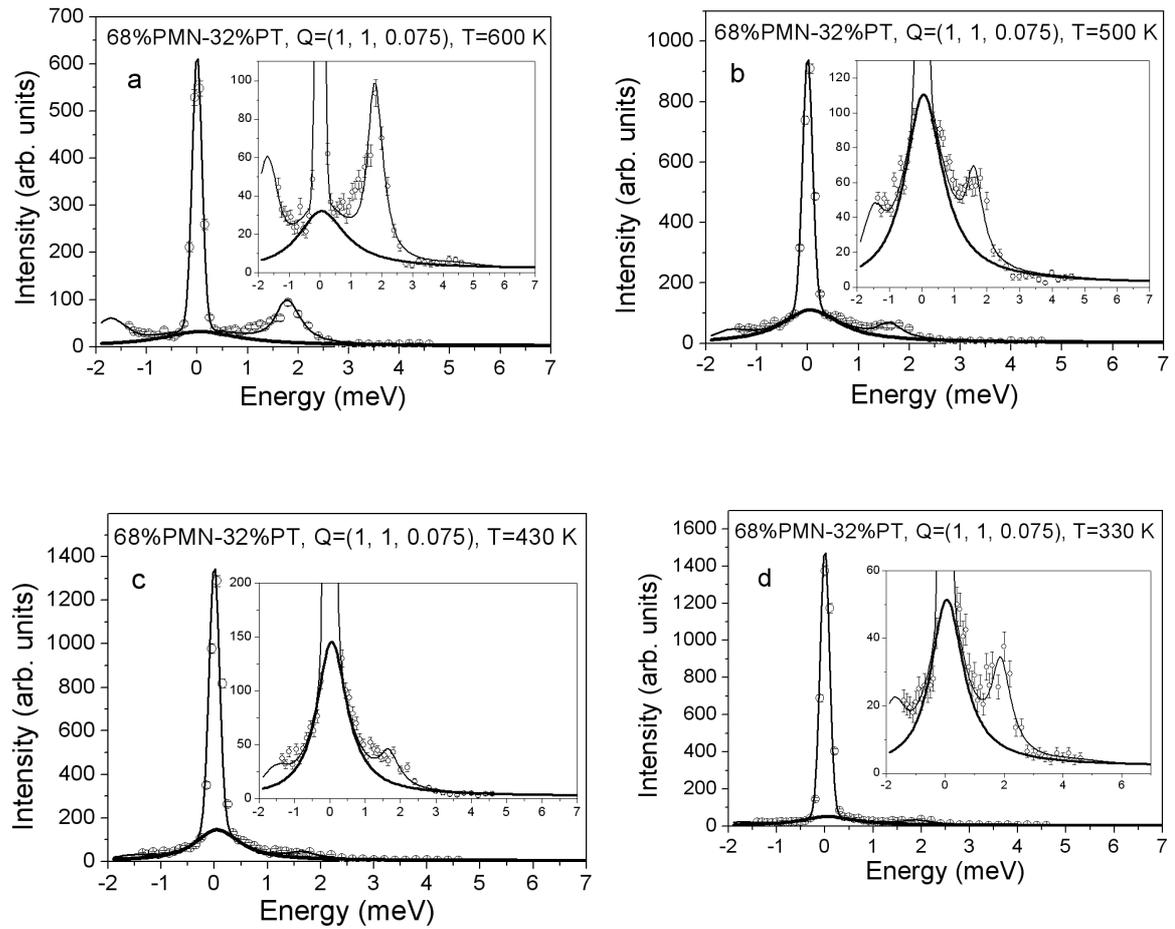

Fig. 7. The temperature of the intensity obtained from experiments in which the wave-vector transfer was held fixed at (1,1,0.75). The fits are described in the text.

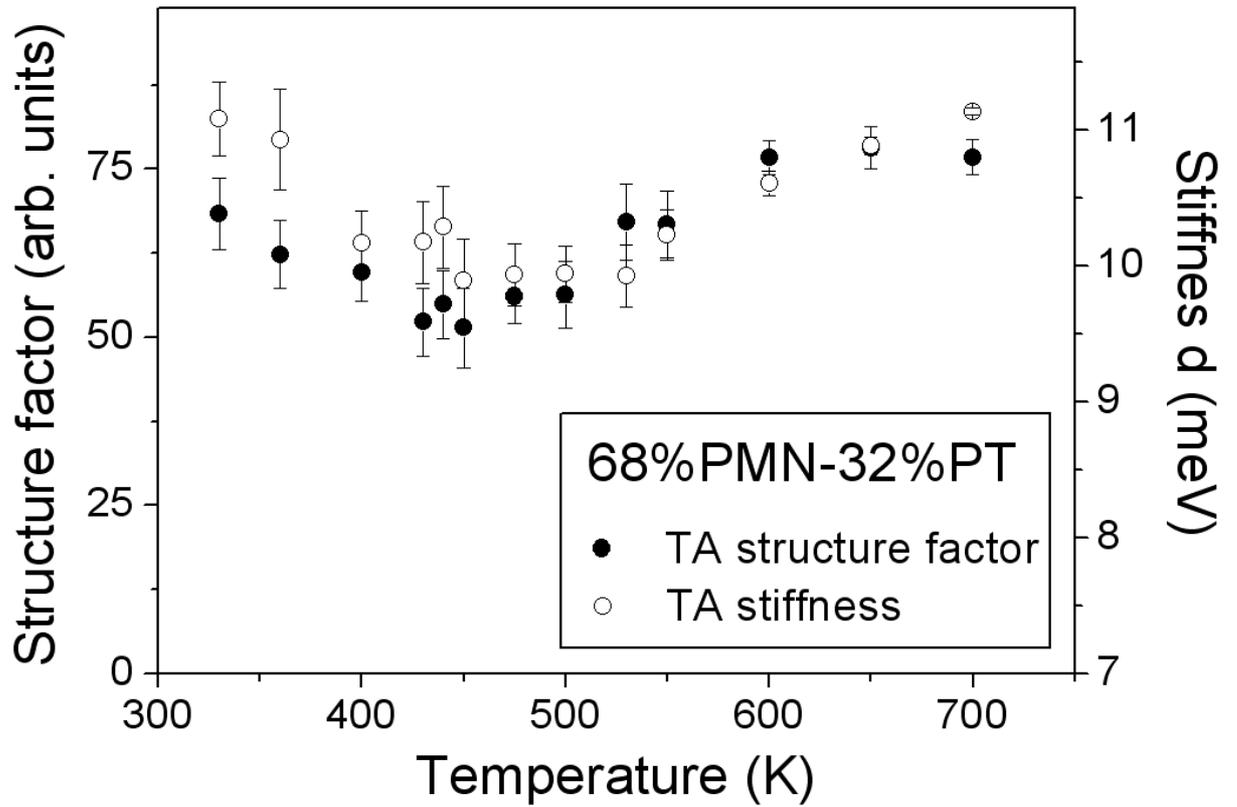

Fig. 8. The temperature dependence of the slope (stiffness) and structure factor of the TA mode as measured near the (1,1,0) Bragg reflection. Note that the corresponding values obtained directly from data taken near the (2,2,0) Bragg reflection show temperature independent behaviour while the fractional changes for the (1,1,0.75) data are also small.

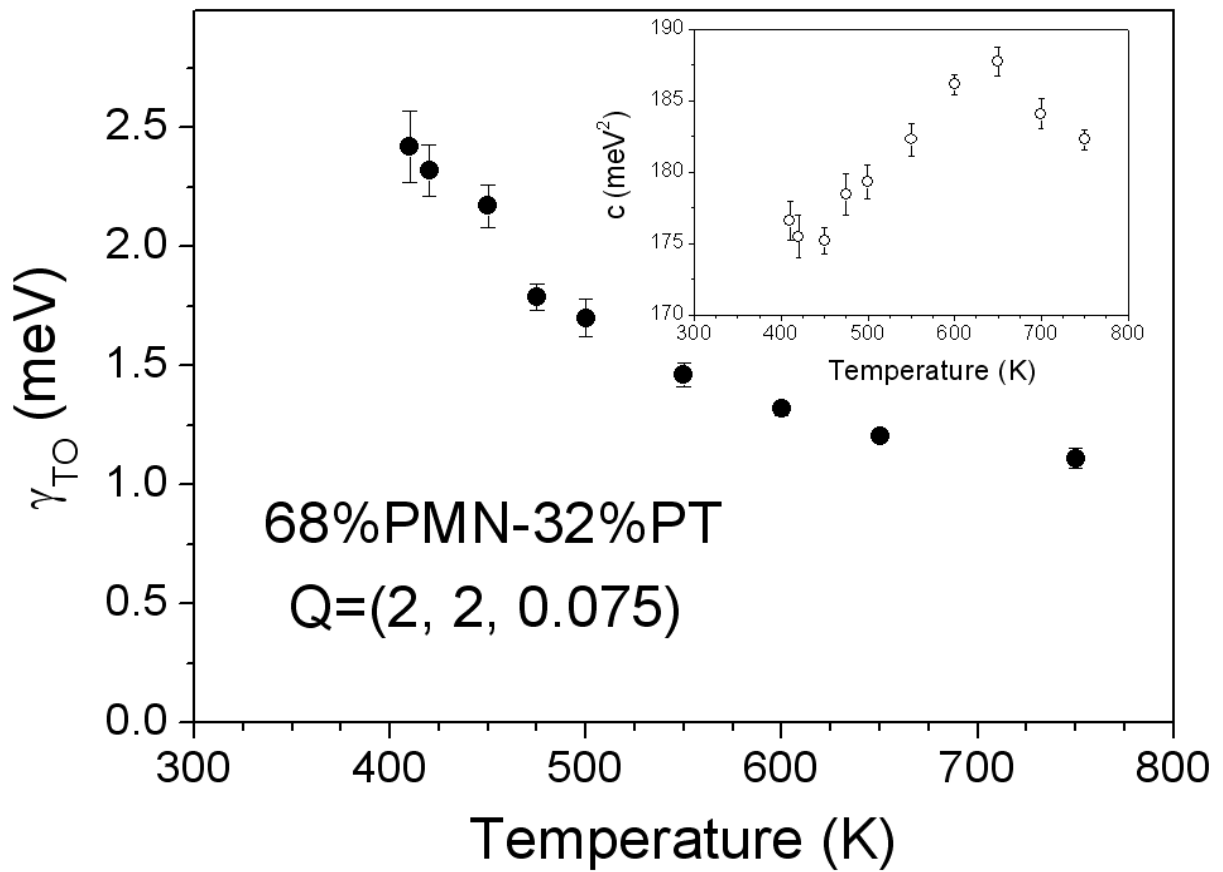

Fig. 9. The temperature dependence of the damping of the TO phonon as measured near the (2,2,0) Bragg reflection. The inset shows the temperature dependence of the stiffness $c$ of the TO phonon as deduced from the data at the (2,2,0.075) position.

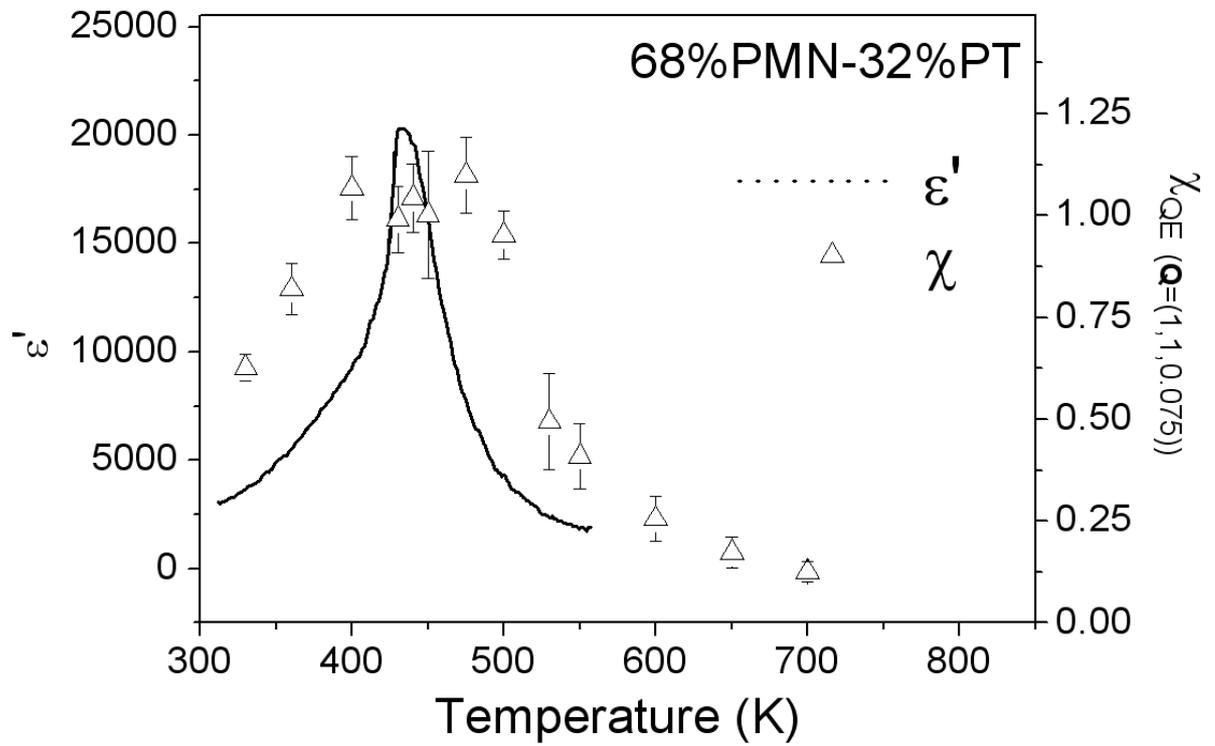

Fig. 10. The temperature dependence of the susceptibility of the quasi-elastic scattering as measured at (1,1,0.75) compared with the dielectric susceptibility [20].

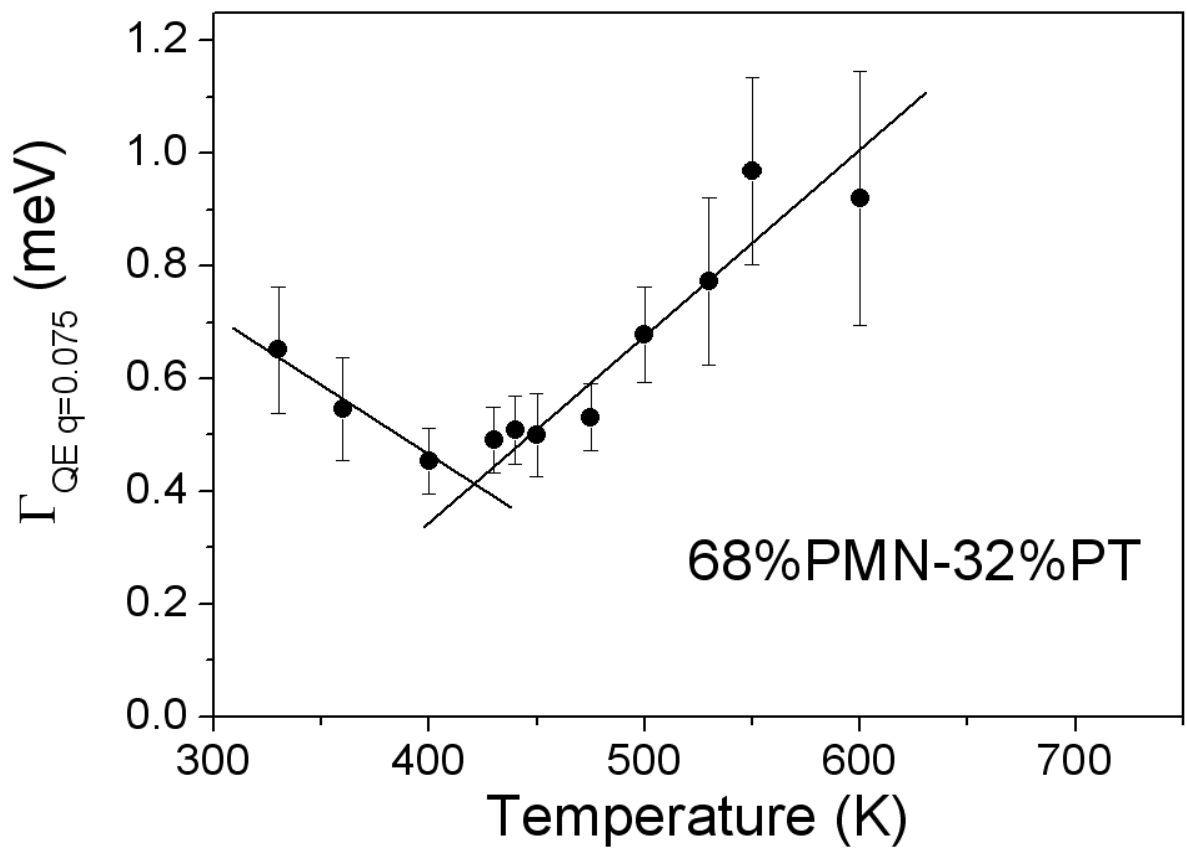

Fig. 11. The temperature dependence of the damping of the quasi-elastic scattering as measured at (1,1,0.75). The fits are to two linear power laws.

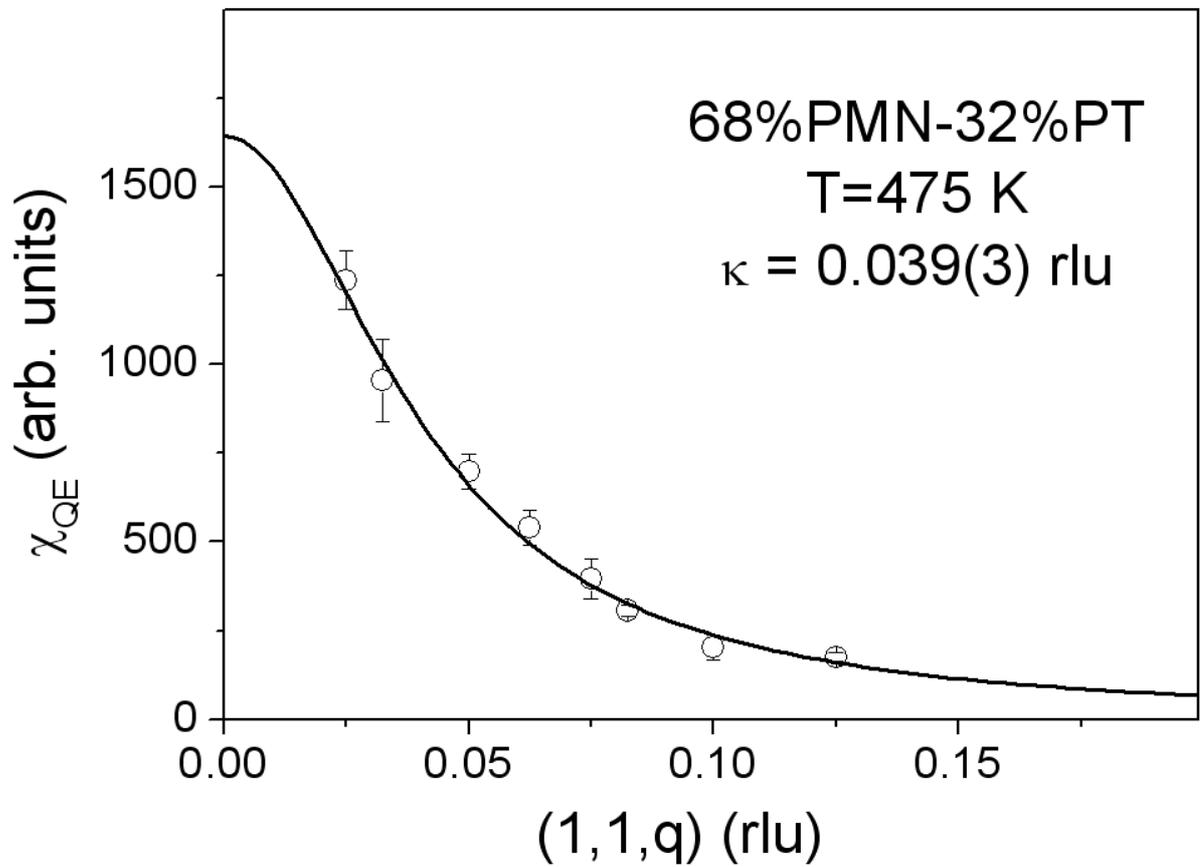

Fig. 12. The wave-vector dependence of the susceptibility of the quasi-elastic scattering measured close to the (110) Bragg reflection.

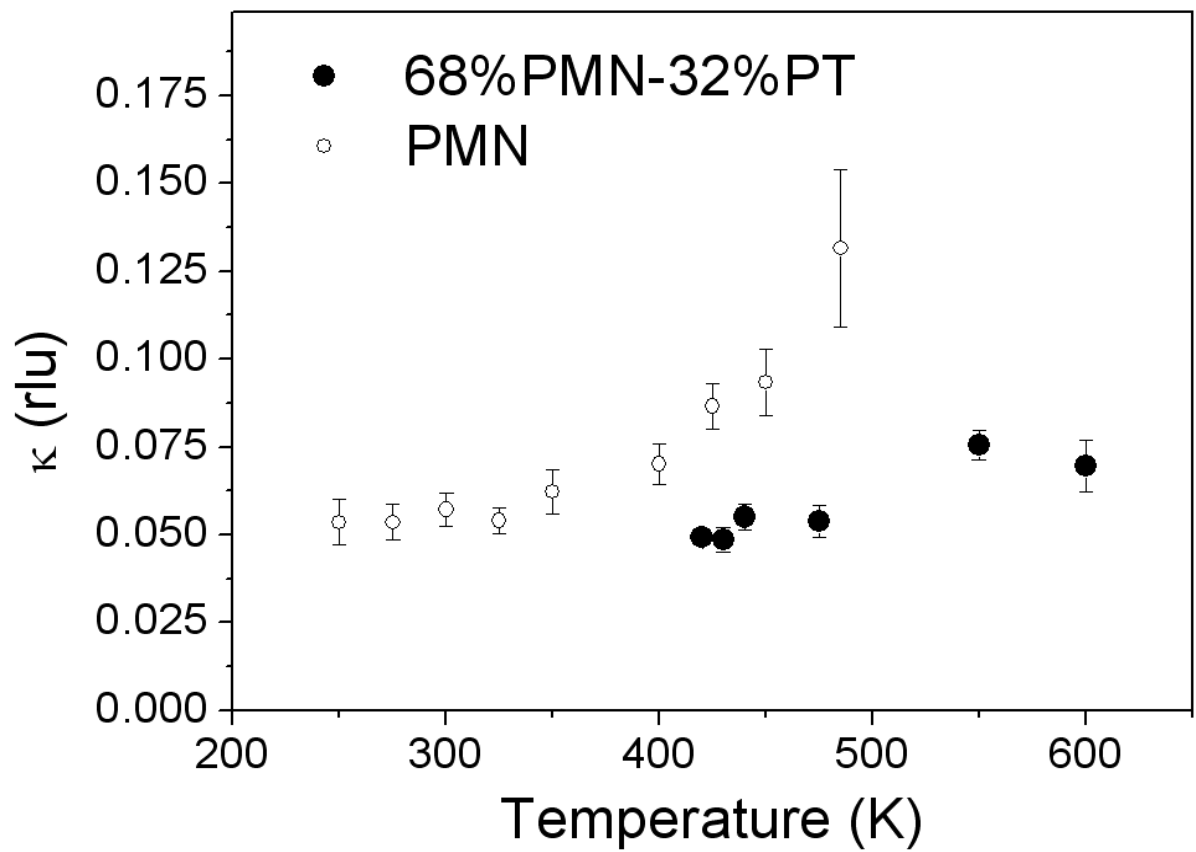

Fig. 13. The width in the reciprocal space of the quasi-elastic scattering as a function of temperature. The measurements were made near the (110) Bragg reflection and the results for 0.68PMN-0.32PT are compared with those from PMN.

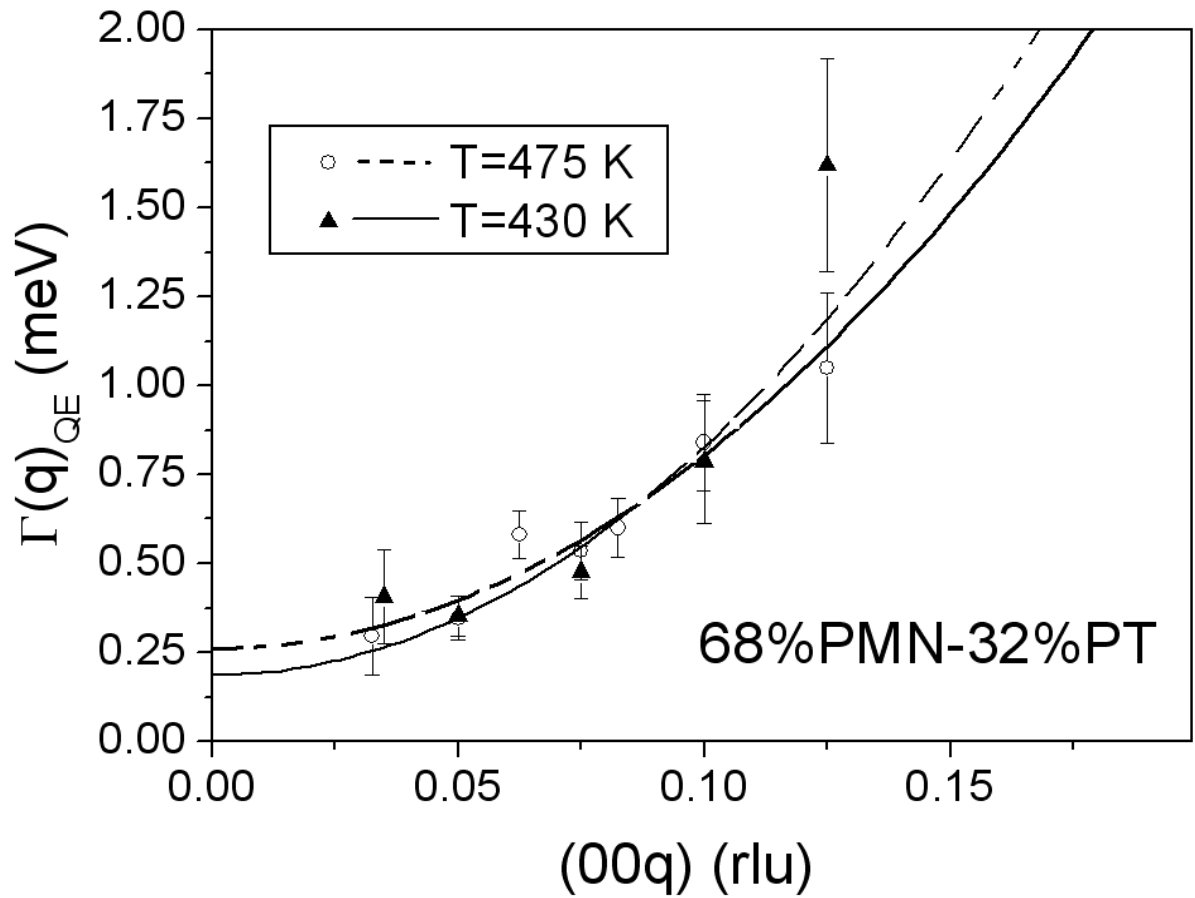

Fig. 14. The wave-vector dependence of the damping of the quasi-elastic scattering measured in the vicinity of the (110) Bragg reflection.

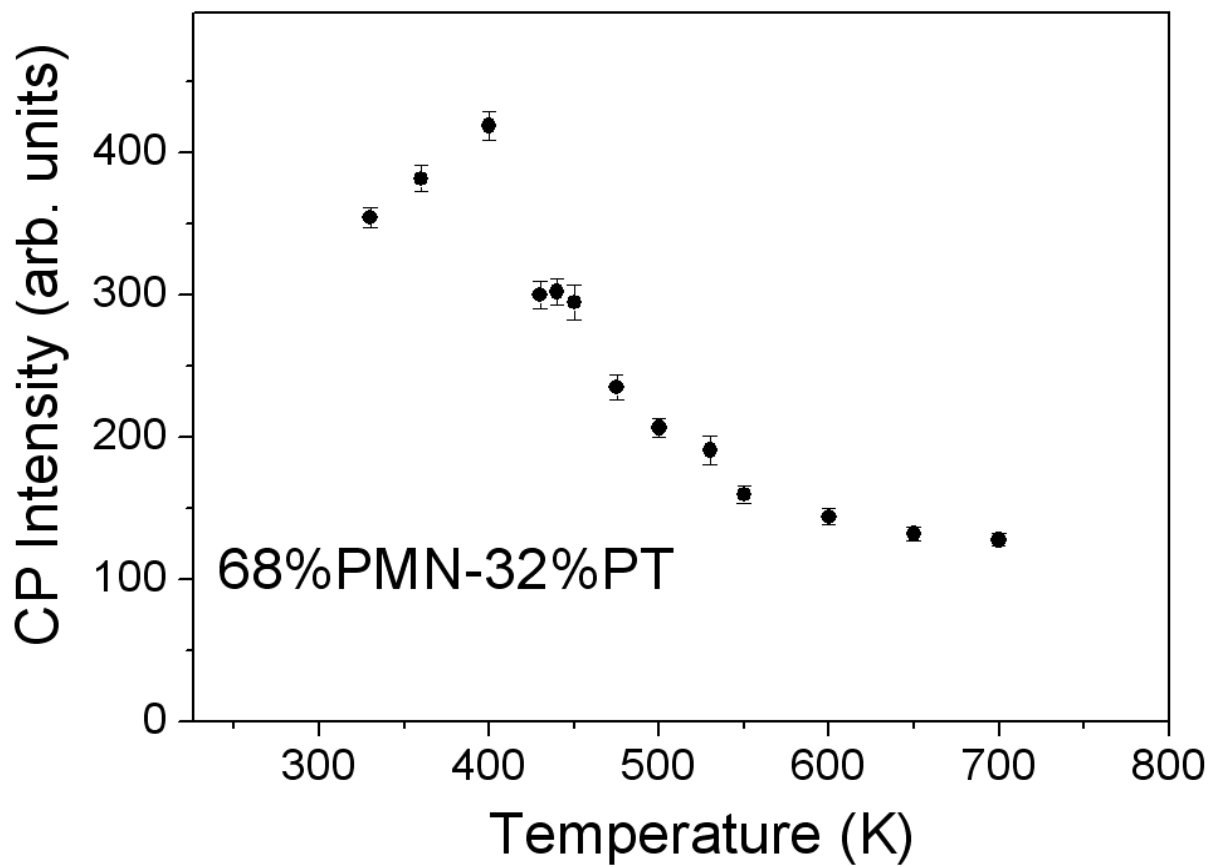

Fig.15. The temperature dependence of the central peak (energy resolution limited peak) as a function of temperature. The results are taken from the data at a wave vector transfer of (1,1,0.75).